\documentclass[aps,prb,twocolumn,groupedaddress,floatfix,eqsecnum,showpacs]{revtex4}

\usepackage[dvips]{graphics,color,floatflt,epsfig} 

\usepackage{amsmath}

\usepackage{amssymb} 

\usepackage{graphicx} 

\usepackage{psfrag} 

\usepackage{bm} 

\input colordvi

\begin{document}

\newcommand{\Sref}[1]{Section~\ref{#1}}

\newcommand{\sref}[1]{Sec.~\ref{#1}}

\newcommand{\tql}{\textquotedblleft} 
\newcommand{\tqr}{\textquotedblright~} 
\newcommand{\tqrc}{\textquotedblright} 

\newcommand{\Refe}[1]{Equation~(\ref{#1})}

\newcommand{\Refes}[1]{Equations~(\ref{#1})}

\newcommand{\fref}[1]{Fig.~\ref{#1}}

\newcommand{\frefs}[1]{Figs.~\ref{#1}}

\newcommand{\Fref}[1]{Figure~\ref{#1}}

\newcommand{\Frefs}[1]{Figures~\ref{#1}}

\newcommand{\reff}[1]{(\ref{#1})}

\newcommand{\refe}[1]{Eq.~(\ref{#1})}

\newcommand{\refes}[1]{Eqs.~(\ref{#1})}

\newcommand{\refi}[1]{Ineq.~(\ref{#1})}

\newcommand{\refis}[1]{Ineqs.~(\ref{#1})}

\newcommand{\PRA }{{ Phys. Rev.} A }

\newcommand{\PRB }{{ Phys. Rev.} B} 

\newcommand{\PRE }{{ Phys. Rev.} E}

\newcommand{\PR}{{ Phys. Rev.}} 

\newcommand{\APL }{{ Appl. Phys. Lett.} }

\newcommand{\PRL}{Phys.\ Rev.\ Lett. }

\newcommand{\OCOM }{{ Opt. Commun.} } 

\newcommand{\JOSA }{{ J. Opt. Soc. Am.} A}

\newcommand{\JOSB }{{ J. Opt. Soc. Am.} A}

\newcommand{\JMO }{{J. Mod. Opt.}}

\newcommand{\RMP}{Rev. \ Mod. \ Phys. }

\newcommand{\etal} {{\em et al.}}

\title{Complete Band Gaps in 2D Photonic Quasicrystals}

\author{Marian Florescu$^{1}$} 

\email[Electronic Address:]{florescu@princeton.edu}

\author{Salvatore Torquato$^{2,3}$}

\author{Paul J. Steinhardt$^{2,3}$}

\affiliation{$^1$ Department of Physics, Princeton University, Princeton, New Jersey,
  08544, USA}

\affiliation{$^2$Department of Chemistry, Princeton University, Princeton, New Jersey
  08544, USA}

\affiliation{$^3$Princeton Center for Theoretical Sciences, Princeton University,
  Princeton, New Jersey 08544, USA}

\date{\today}

\begin{abstract} 
We introduce a novel optimization method to design the first examples of photonic
quasicrystals with substantial, complete photonic band gaps (PBGs): that is, a range of
frequencies over which electromagnetic wave propagation is forbidden for all directions
and polarizations.  The method can be applied to photonic quasicrystals with arbitrary
rotational symmetry; here, we illustrate the results for 5- and 8-fold symmetric
quasicrystals. The optimized band gaps are highly isotropic, which may offer advantages
over photonic crystals for certain applications.
\end{abstract}

\maketitle

\section{Introduction} 
\label{secI}
Photonic materials with complete photonic band gaps (PBGs), i.e., frequency ranges over
which electromagnetic wave propagation is prohibited for all directions and polarizations,
are artificial dielectric heterostructures that enable control of the generation and flow
of light.  Their development \cite{c1,c2} has evolved dramatically in the last decade and
their unusual properties have found a wide range of applications including efficient
radiation sources,\cite{radSources} telecommunications devices ,\cite{telecom} sensors
\cite{PCsensor} and optical computer chips.\cite{PCcompChips} Until recently, the only
known materials with complete PBGs were photonic crystals comprised of a periodic
arrangements of dielectric materials.  In photonic crystals, the opening of the PBG is
understood to be governed by the synergetic interplay between Bragg scattering resonances
of the periodic dielectric array and the Mie resonances of individual dielectric
scattering centers.\cite{esaj}

\begin{figure}[h]
{\centerline{\includegraphics*[width=0.75\linewidth]{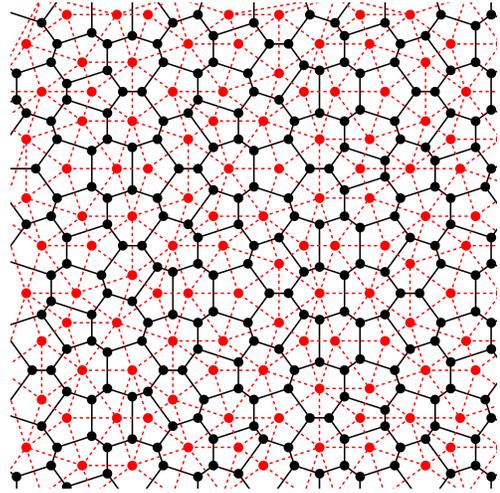}}}
\caption{Protocol described in Sec.~\ref{secII} maps the vertices of a rhombus tiling
  (red/grey points) into a network of cells whose vertices are trivalent (black segments
  and points). To construct a photonic material with a complete band gap, edges are
  replaced by a wall of dielectric of finite thickness and the vertices are replaced by
  cylinders of finite radius. }\label{fig1}
\end{figure}

A similar combination of scattering mechanisms is possible in photonic quasicrystals, in
which the dielectric materials are arranged in a pattern with long-range quasiperiodic
translational order and rotational symmetries forbidden to crystals (such as 5-fold
symmetry in two dimensions and icosahedral symmetry in three dimensions).\cite{stein}
Quasiperiodic order also produces Bragg scattering \cite{stein}.  In fact, there have been
numerous studies of photonic quasicrystals with PBGs in the
literature.\cite{qpc1,qpc2,qpc3,qpc4,qpc5,qpc6,qpc7,qpc8} Their band gaps are found to be
considerably more isotropic and, for two-dimensional quasicrystals, the dielectric
contrast required to open the band gaps for TM polarization (electric field oscillating
out of the plane) is smaller than the contrast required for the periodic
counterparts.\cite{qpc2} Most of the photonic quasicrystalline structures considered only
have band gaps for either TM or TE polarized radiation.\cite{qpc1,qpc5,qpc7} Two examples
of complete band gaps have been found previously;\cite{qpc4,qpc8} here we present a
systematic method to produce substantially greater fundamental complete band gaps for
photonic quasicrystals with arbitrary symmetry.

In general, the architectures required for optimal TM band gaps are quite different than
optimal architectures for TE polarization (electric field oscillating in the plane).
Structures based on a distribution of isolated dielectric inclusions ({\it e.g.}, rods) is
best for TM band baps, while structures based on connected dielectric networks are optimal
for TE.\cite{tmteLoc} For the case of quasicrystalline systems, with their rich variety of
local environments, the problem of finding an optimal compromise becomes more
difficult. Consequently, until now, there has not been a systematic procedure for
designing quasiperiodic photonic structures with sizable PBGs for all directions of
propagation and polarizations.

\begin{figure*}[t]
{\centerline{\includegraphics*[width=0.75\linewidth]{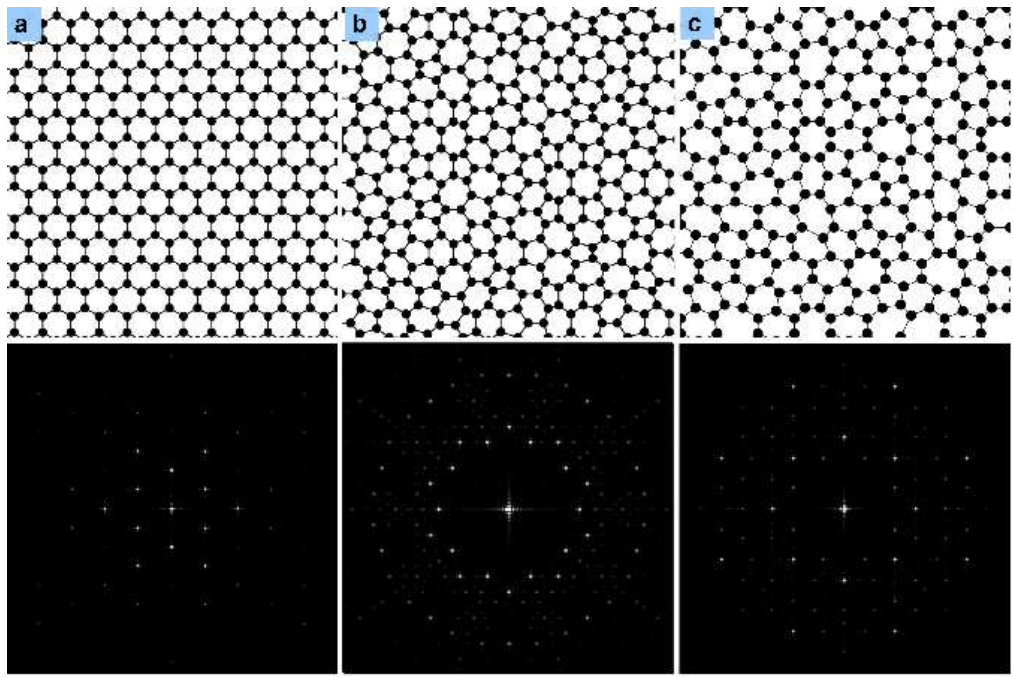}}}
\caption{Optimal photonic crystal and quasicrystal heterostructures and their diffraction
  patterns derived from: (a) a periodic 6-fold symmetric point pattern; (b) a 5-fold
  symmetric quasicrystal (Penrose) tiling; (c) an 8-fold symmetric quasicrystal
  tiling.}\label{fig2}
\end{figure*}

In this paper, we introduce a novel optimization scheme discussed in
Ref.~\onlinecite{ref1} (where it was applied to disordered heterostructures) to design
two-dimensional photonic quasicrystals with substantial complete PBGs, comparable to the
largest band gaps found for photonic crystals.  The examples illustrated here are based on
the vertices of a Penrose tiling,\cite{penrose} a 5-fold symmetric pattern composed of
obtuse and acute rhombi and on the vertices of an octagonal tiling, \cite{octa} an 8-fold
symmetric pattern composed of squares and rhombi.

In Section~\ref{secII}, we describe how the quasiperiodic patterns are constructed. In
Section~\ref{secIII}, we present the results of the band structure calculations and
analyze the properties of the resulting PBGs including the field distribution of the
photonic modes that define the photonic band edges. Section~\ref{secIV} provides
concluding remarks.

\begin{figure}[ht]
\begin{center}
\includegraphics*[width=1\linewidth]{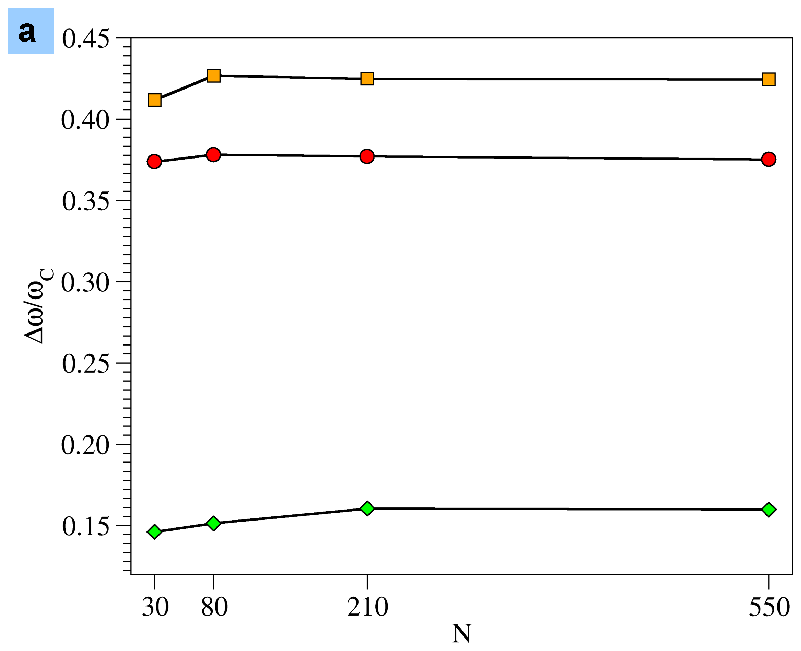}
\includegraphics*[width=1\linewidth]{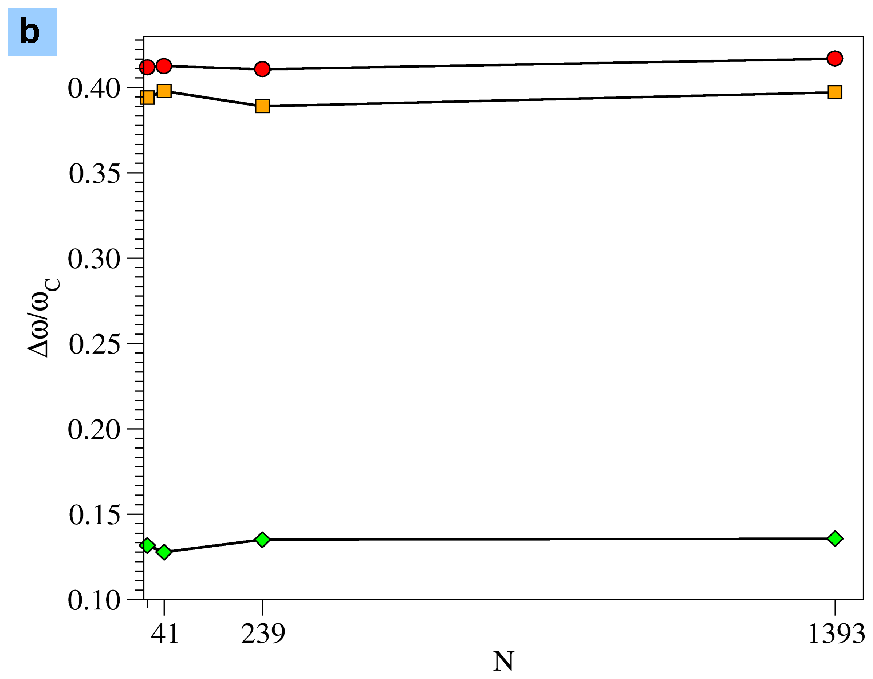}
\end{center}
\caption{ The fractional PBG width for (a) 5-fold symmetric and (b) 8-fold symmetric
  photonic quasicrystal as a function of the number of particles in the periodic
  approximant for the optimal TM (red circles), TE (orange squares) and complete (green
  diamonds) band gaps.  }\label{fig3}
\end{figure}

\section{Nearly Optimal Photonic Band Bap Structures from  Quasiperiodic Point Patterns}
\label{secII}
The Penrose and octagonal quasicrystalline point patterns considered in this work are
constructed by the projection and cut method,\cite{brujin} which consists of projecting
the points of an $n$-dimensional hypercubic lattice into the three dimensional space,
where $n=4$ for the octagonal lattice and $n=5$ for the Penrose lattice. In general, the
lack of spatial periodicity makes it impossible to define a unit cell for band gap
calculations; we address this by using a series of rational periodic approximants of the
quasicrystal. In the case of the Penrose lattice periodic approximants are constructed by
the same projection method by replacing by a Fibonacci ratio ($\tau_n=p_n/q_n={1/1, 2/1,
  3/2, 5/3, ...}$). The unit cell is rectangularly shaped and its area grows as n
increases and the rational approximant approaches $\tau$. In particular, for a rational
approximant $\tau_n=p_n/q_n$, the unit cell has dimensions $L_x=5(p_n+q_n(\tau-1), L_y =
\sqrt{3-\tau}(p_n\tau+q_n)$ and contains $N=10p_n(p+2q_n)$ vertices of the rhombic
pattern.  In the case of the octagonal quasilattices, we replace by its continuous
fraction series ($\rho_n=p_n/q_n={1/1, 3/2, 7/5,...}$), and the resulting periodic
approximant has a square shaped unit cell of side $L=p_n+\rho q_n$ containing
$N=(p_n+(\rho^2+\rho)q_n) (p_n+(\rho^2-\rho)q_n)$ vertices of the square-rhombic
pattern. The periodic approximants constructed this way results in tilings consisting of
identical tiles as the original quasiperiodic tiling and transition from quasiperiodicity
to periodicity is accommodated through the flipping of a certain number of tiles. These
rational periodic approximants are known to be the best approximants of the quasiperiodic
lattice as they achieve the selected periodicity by introducing the minimum density of
defects with respect to the perfect quasiperiodic tiling.\cite{RaQpc} The vertices of the
tilings obtained by this method form a sequence of quasicrystal approximant point pattern.

The next step is to find the arrangement of dielectric material around the point pattern
that produces the optimal TM, TE and complete PBGs. Identifying these optimal dielectric
distributions is well known to be a daunting computational task, despite the recent
development of rigorous mathematical optimization methods.\cite{c11,c12,c13}

For the case of TM radiation only, a nearly optimal decoration for a given point pattern
is obtained by placing identical dielectric cylinders centered at each point and adjusting
the radius.  For the optimal radius, there is a coincidence of Mie and Bragg scattering
effects that lead to substantial band gaps.\cite{esaj} The dielectric cylinders support
Mie scattering resonances and, for frequencies above the lowest order resonance, the
scattered radiation is out of phase with the incident wave.  This destructive interference
prohibits radiation propagation and favors the opening of a complete TM gap.\cite{qpc7} At
the same time, the quasiperiodic long range order in quasicrystalline systems results in a
dense (no minimum separation) collection of Bragg peaks; however, it is notable that most
of Bragg peak intensities are infinitesimally small. The first set of intense Bragg peaks
is associated with scattering on planes separated by the average distance in the
quasicrystalline point pattern and can be employed to define an effective Brillouin
zone. Therefore, as for photonic crystals, Bragg scattering also contributes to the PBG
formation in photonic quasicrystals such that, whenever the wavevector of the incident
radiation is directed along the effective Brillouin zone boundary, the reflected and
refracted waves interfere to cancel the incoming wave and prevent its propagation inside
the structure. An optimal band gap occurs when the two scattering mechanism reinforce each
other on a given spectral range.  This approach can be applied to find near optimal TM
band gap for the 5- and 8-fold quasicrystal patterns obtained by projection.\cite{c13}

To obtain the optimal PBGs for TE radiation, where the electric field is oriented in the
plane of the scatters, necessitates a different dielectric arrangement. Instead of
isolated cylinders, a connected network of dielectric with air pockets in between is
favored.  For example, a commonly used configuration for photonic crystals is \tql
inverted\tqr compared to the optimal structures for TM PBGs, i.e., placing an identical
air cylinder at each point so as to produce a connected network of dielectric material.
However, in the case of quasicrystalline structures, we find this method fails to produce
sizable fundamental TE PBGs.  The main reason is that the inverted structure has a very
non-uniform distribution of dielectric scattering regions that broadens the distribution
of resonances. Similarly, placing walls along the edges of each Penrose tile as in
Ref.~\onlinecite{qpc1} has rather non-uniform connectivity and fails to produce an optimal
TE PBG.

To overcome this problem, we employ a protocol introduced in Ref.~\onlinecite{ref1} that
converts a general point pattern into photonic heterostructures with sizeable, nearly
optimal TE PBGs.  We briefly review the protocol here: Consider an arbitrary point pattern
denoted by red disks in \fref{fig1}. In the example in the figure, the point pattern is
enclosed in a square domain with periodic boundary conditions. First, we perform a
Delaunay tessellation of the original point pattern. This provides a triangular
partitioning that minimizes the standard deviation of the triangle angles around the mean
of $\pi/6$.  Next, associated with each triangle is a centroidal point; these are
connected with line segments to form cells around the original points.  The nearly optimal
heterostructure for TE band gaps is obtained by decorating the edges of the cell network
with walls (along the azimuthal direction) of dielectric material of uniform width $w$, as
displayed in \fref{fig1}.  We note that this procedure is universal and can be used to
generate connected network architectures based on periodic, quasiperiodic and disordered
point patterns. Finally, the optimal structure is obtained by varying a single parameter,
the cell thickness, and identifying the value that maximizes the TE band gap.

The total number of cells in the network is the same as the number of points in the
original point pattern and each vertex of the network is trivalent.  The cells can be
regarded as individual scattering objects supporting electromagnetic resonances, which
become coupled to each other when placed into the connected network.

With an additional step, the protocol can be used to a heterostructure with a nearly
optimal {\it complete} PBG, i.e., maximal overlap between TM and TE gaps. Namely, each
vertex of the trihedral network constructed above is decorated with a cylinder or radius
$r$.  The number of cylinder is equal to twice the number of cells. The optimal complete
band gap obtained by varying the two free parameters, the cell wall thickness $w$ and
cylinder radius $r$. As shown in \fref{fig2}, the Fourier spectrum of the decorated
photonic structures obtained through this protocol fully preserve the symmetries of the
original point pattern. In comparison, alternative designs based on \tql inverted\tqr
architectures or decorations of the quasicrystal tile edges with dielectric walls and its
vertices with dielectric cylinders fails to open sizable complete PBGs.

\begin{figure}[t!]
\begin{center}
\includegraphics*[width=1\linewidth]{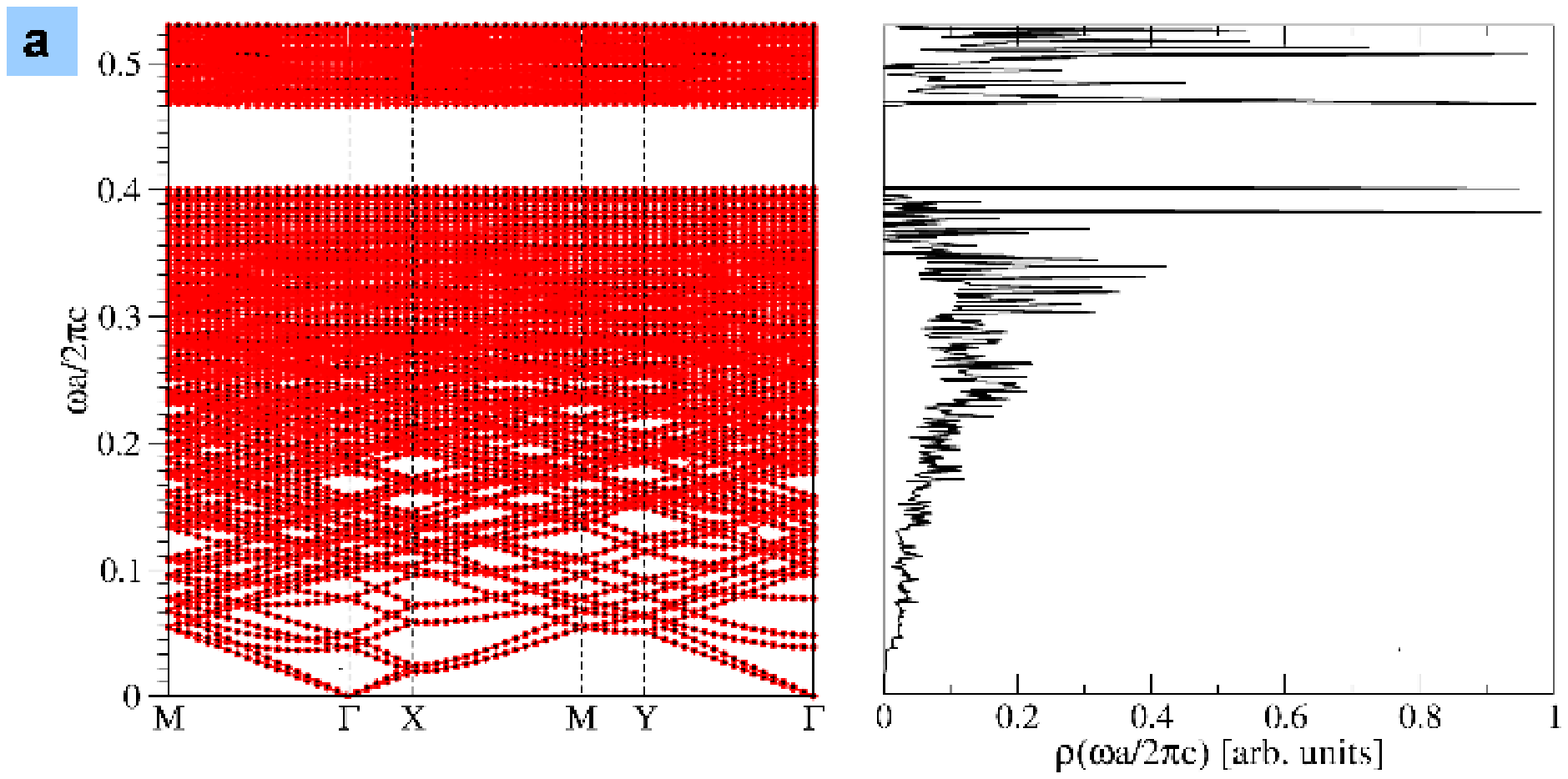}
\includegraphics*[width=1\linewidth]{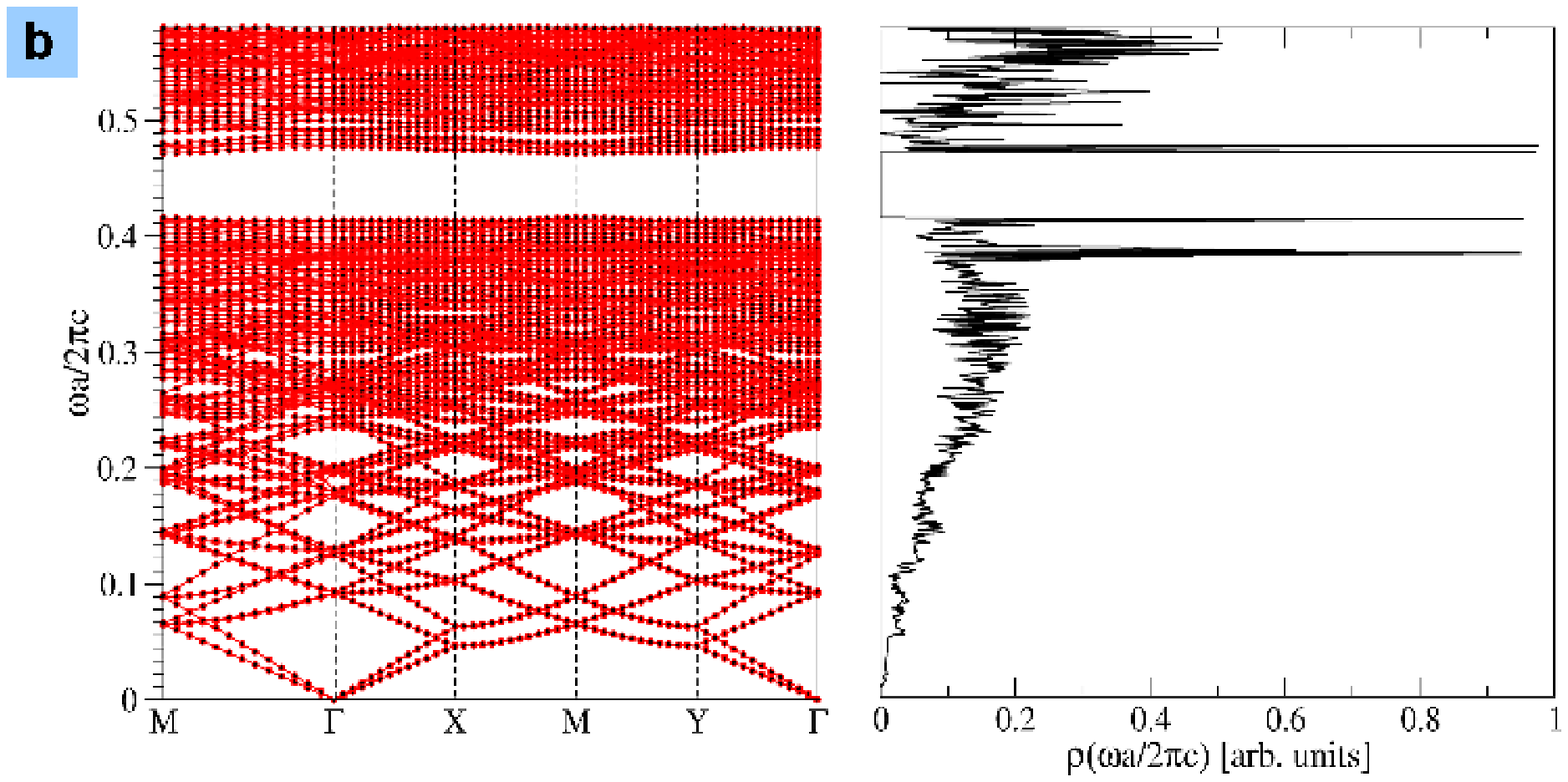}
\end{center}
\caption{Photonic band structure and density of states for 5-fold and 8-fold symmetric
  photonic quasicrystals. (a) Photonic band structure and the corresponding density of
  states for a network constructed via the protocol in Sec.~\ref{secII} from a 2/1
  periodic approximant of the Penrose tiling with optimized cylinder radius $r/a=0.157$
  and wall thickness $w/a=0.042$. The structure has a complete PBG of 15.14\% with central
  frequency $\omega c/(2\pi a)=0.43$. (b) Photonic band structure and the corresponding
  density of states for a network constructed from a 3/2 periodic approximant of the
  octagonal quasilattice with optimized cylinder radius $r/a$=0.1613 and wall thickness
  $w/a=0.0136$. The structure has a complete PBG of 13\% with central frequency $\omega
  c/(2\pi a)=0.44$.  }\label{fig6}
\end{figure}

\section{Photonic Band Gap Results}
\label{secIII}
We compute the photonic band structures for the optimal heterostructures (described in
detail below) using a conventional plane-wave expansion formalism.\cite{Johnson} For this
approach, which assumes a periodic structure, we use a sequence of increasingly accurate
periodic approximants and check for convergence.  For the purposes of illustration, we
assume the photonic materials are composed of silicon (with dielectric constant $\epsilon=
11.56$) and air.  In all the numerical simulations, we use $32^2\times N_P$ plane waves to
achieve convergence accuracy of better than 1\% for the lowest $N_P$ photonic bands.  Here
$N_P$ represents the number of scattering centers (TM case), cells (TE case) or centers
and cells (TM+TE case). Most of the band structure calculations are preformed around a
contour along the first Brillouin zone of the respective system, which includes the high
symmetry k-space points, $\mathbf{\Gamma}=0$, $\mathbf{X}= \mathbf{b}_1/2 $, $\mathbf{M}=(
\mathbf{b}_1 +\mathbf{b}_2 )/2$, and $\mathbf{Y}= \mathbf{b}_2/2$, where $ \mathbf{b}_1$
and $\mathbf{b}_2$ are the basis vectors of the reciprocal lattice considered.

\noindent
{\it TM Band Gaps:} To obtain optimal TM PBGs for a given symmetry, dielectric cylinders
are placed at the vertices of the tilings obtained by projection.

For example, an optimized 5-fold symmetric quasicrystalline pattern derived using the
protocol from a $n=5/3$ periodic approximant of the Penrose pattern displays a TM PBG of
$\Delta \omega/\omega_C =39\%$.  An analogous optimized structure based on the $n=7/5$
periodic approximant of the octagonal lattice displays a TM PBG of $\Delta
\omega/\omega_C=42\%$, where $\omega_C$ is the gap central frequency. The optimal radius
of the cylindrical inclusion is $r/a=0.177$ for the Penrose tiling and $r/a=0.189$ for the
octagonal lattice, were $a$ is the side length of the rhombic tiles obtained by
projection.  As a check, we perform a convergence analysis as the order of the periodic
approximant increases; as shown in \fref{fig3}, the results are fully convergent.

We find that the band gap for the optimal structures always occurs between bands $N_P$ and
$N_P+1$ where $N_P$ is the number of points in the periodic approximant point pattern.
This is an analogous behavior to the band folding that occurs in periodic structures when
analyzed with a super-cell approach.

\noindent
{\it TE Band Gaps:} The optimal TE band gap designs are obtained by the protocol described
in Sec.\ref{secII}.  They correspond to the network of walls connecting the centroids of
the Delaunay tiling.  In the case of the 5-fold symmetric quasicrystal, the optimum has
wall width $w/a = 0.103$; the TE band gap is $\Delta \omega/\omega_C=42.3\%$, the largest
ever reported for a photonic quasicrystal.  For the octagonal quasicrystal, the optimal
structure has dielectric material width $w/a = 0.106$ along each edge, which produces a TE
band gap of $\Delta \omega/\omega_C=39.2\%$.

The TE band gap formation is analogous to the TM case in that it involves an interplay
between scattering from individual cells and the Bragg scattering of the quasiperiodic
arrangement of scattering planes.  Similar to the case of TM polarized radiation, we also
obtain that the band gap for the optimal structures always occurs between bands $N_P$ and
$N_P+1$ where $N_P$ is now the number of cells in the structure.

\clearpage
\begin{figure}[h!]
{\centerline{\includegraphics*[width=1\linewidth]{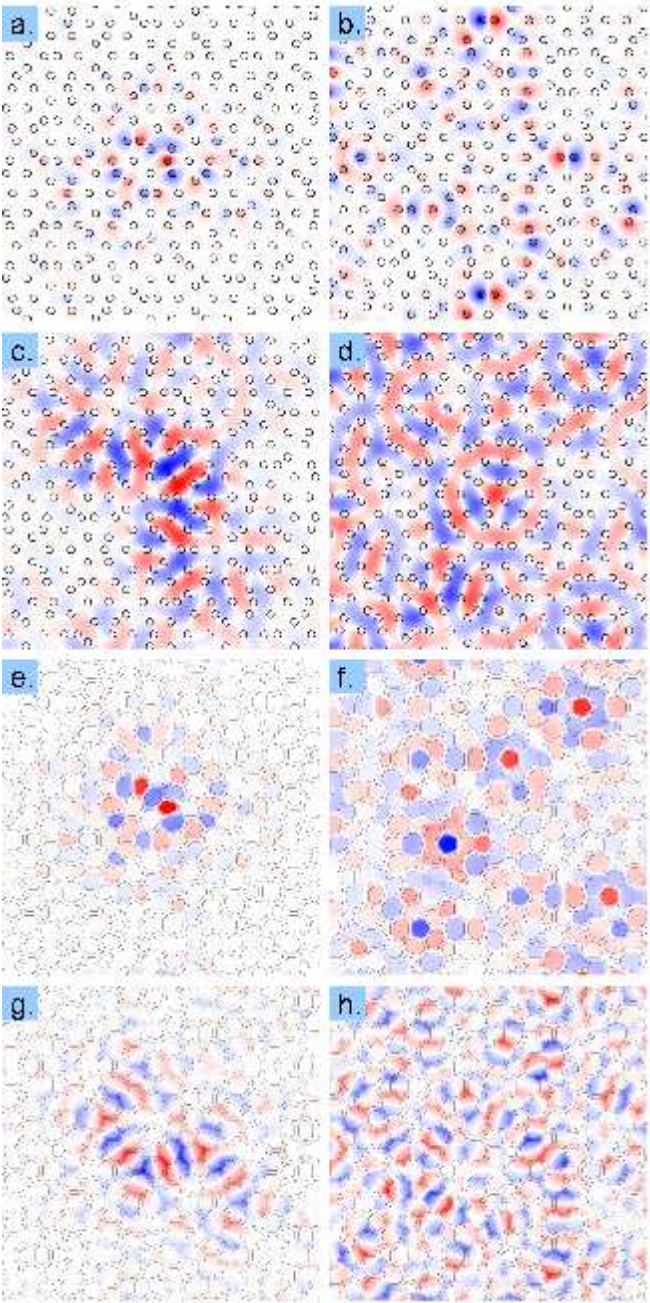}}}
\caption{ Panels (a)-(d) show the electric field distribution for a 5-fold symmetric
  photonic quasicrystal optimized for TM polarized radiation. The structure consists of
  dielectric cylinders of radius $r/a=0.177$ placed at the vertices of a Penrose tiling
  and displays a TM photonic band gap of 36.5\%. Lower (a) and upper (c) band edge modes
  display a well-defined degree of localization; modes just below the lower band edge (b)
  and just above the upper band edge (d) display an extended character. Panels (e)-(h)
  show the magnetic field distribution in 5-fold symmetric quasicrystalline network
  optimized via the protocol in Sec.~\ref{secII} for TE polarized radiation. The structure
  consists of trihedral network with wall thickness $w/a=0.102$ and displays a TE PBG of
  42.5\%. The lower (e) and upper (g) band edge modes display a high degree of spatial
  concentration and modes just below the lower band edge (f) and just above the upper band
  edge (h) display an extended character. }\label{fig4}
\end{figure}

\begin{figure}[h!]
{\centerline{\includegraphics*[width=1\linewidth]{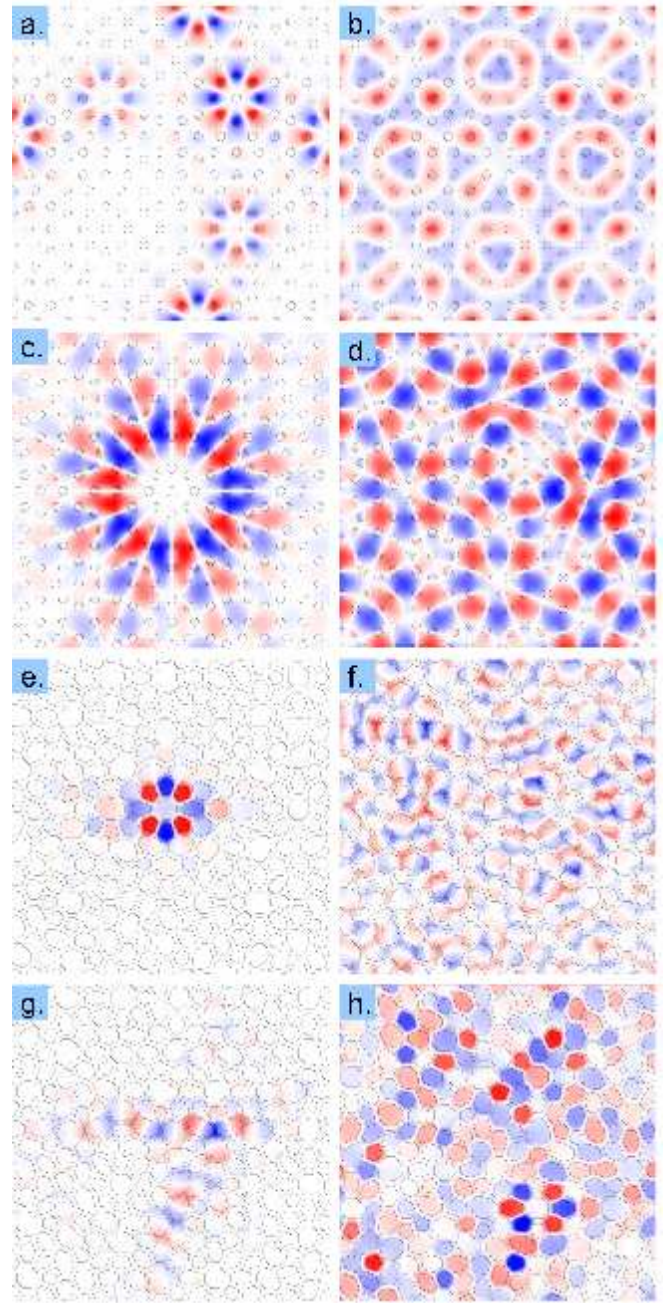}}}
\caption{ Panels (a)-(d) show the electric field distribution in 8-fold symmetric
  structures for TM polarized radiation. The structure consists of dielectric cylinders of
  radius $r/a=0.189$ and displays a TM PBG of 42\%. Lower (a) and upper (c) band edge
  modes displaying a well defined degree of localization; modes just below the lower band
  edge (a) and just above the upper band edge (d) displaying an extended character. Panels
  (e)-(h) show analogous magnetic field distributions in 8-fold symmetric heterostructures
  optimized via the protocol in Sec.~\ref{secII}for TE polarized radiation. The structure
  consists of trihedral network with wall thickness $w/a=0.101$ and displays a TE PBG of
  40\%. Lower (e) and upper (g) band edge modes display a well-defined degree of spatial
  concentration; modes just below the lower band edge (f) and just above the upper band
  edge (h) display an extended character.}\label{fig5}
\end{figure}
\clearpage

\noindent
{\it Complete Band Gap:} For the complete band gap, we find that the optimal structure
consists of placing a wall along each edge and a cylinder at each trihedral vertex of the
network generated by the protocol in Sec.~\ref{secII}.  The optimum for the 5-fold
symmetric case has cylinder radius is $r/a =0.157$ and cell wall thickness $w/a=0.042$.
For the 8-fold symmetric point patterns, the optimum is $r/a=0.167$ and $w/a=0.014$. The
scattering properties of the individual scattering centers and cells are again essential
in the band gap opening and the complete band gap occurs always between bands $3 N_P$ and
$3 N_P+1$, where $N_P$ is the number of points in the periodic approximant point pattern
(there are $3 N_P$ total scattering units in the system, $2 N_P$ dielectric disks and
$N_P$ dielectric cells).

The resulting optimal 5-fold symmetric structure displays complete (TM and TE) PBG of
16.5\% --- the first complete band gap ever reported for a photonic quasicrystal with
5-fold symmetry and comparable to the largest band gap (20\%) found for photonic crystals
with the same dielectric contrast.\cite{hexFullGap} The optimal 8-fold symmetric structure
has a full PBG of 13.5\%.

As shown in \fref{fig6}, the band gap size obtained by considering a contour along the
Brillouin zone is completely consistent with the density of states calculations.  Similar
to the case of the TM and TE PBGs, the system-size study presented in \fref{fig3} reveals
that the band gap size becomes independent of the size of the periodic approximant when
$N_P\ge 100$ (with $N_P$ the number of points in the original point pattern), implying
that our results have converged to the large-system limit.

For a photonic crystal, the states at the photonic band edges are propagating modes such
that the electromagnetic field is distributed throughout the system. If the periodicity is
disturbed, localized states begin to fill in the gap so that the states just below and
just above become localized. Although the nature of the electromagnetic states in
quasiperiodic structures needs to be further investigated, similar features are displayed
by photonic quasicrystalline systems. In \frefs{fig4} (a)-(d) and \ref{fig5} (a)-(d), we
compare the azimuthal electric field distribution for modes below or above the band gap
(plots (b) and (d) in \frefs{fig4} and \ref{fig5}), which display an extended character
with the field distributed among many sites; and then modes at the band edges (plots (a)
and (e) in \frefs{fig4} and \ref{fig5} ), which show a high degree of spatial
concentration.  We also find that the formation of the TM band gaps is closely related to
the formation of electromagnetic resonances localized within the dielectric cylinders (as
shown by the (a) and (b) plots in \frefs{fig4} and \ref{fig5}) and that there is a strong
correlation between the scattering properties of the individual scatterers (dielectric
cylinders) and the band gap location. In particular, the largest TM gap occurs when the
frequency of the first Mie resonance coincides with the lower edge of the photonic band
gap.\cite{qpc7} Analogous to the case of periodic systems, the electric field for the
lower band-edge states is well localized in the cylinders (the high dielectric component),
thereby lowering their frequencies; and the electric field for the upper band-edge states
are localized in the air fraction, increasing their frequencies. As shown in \frefs{fig4}
(e)-(h) and \ref{fig5} (e)-(h), an analogous behavior occurs for the azimuthal magnetic
field distribution for TE modes: for states near the lower edge of the gap, the azimuthal
magnetic field is mostly localized inside the air fraction and presents nodal planes that
pass through the high index of refraction fraction of the structure, while states near the
upper edge display the opposite behavior.

\section{Conclusions}
\label{secIV}
In sum, we have shown that it is possible to expand the spectrum of dielectric materials
with sizeable complete band gaps to include photonic quasicrystals by introducing a novel
constrained optimization procedure.  The quasicrystalline photonic structures can be
manufactured using standard fabrication techniques used for photonic
crystals.\cite{hexFullGap}

Photonic crystals, the quasicrystals considered here, and the disordered heterostructures
discussed in Ref. \onlinecite{ref1} are all decorations of hyperuniform
patterns.\cite{hyperunif} A point pattern is hyperuniform if the number variance within a
spherical sampling window of radius $R$ (in $d$ dimensions) grows more slowly than the
window volume for large $R$, i.e., $\langle N^2_R\rangle -\langle N_R\rangle^2= AR^p$,
where $p<d$. Two dimensional crystal and quasicrystal patterns both correspond to $p=1$.
Our conjecture in Ref.~\onlinecite{ref1}, which was supported by a variety of examples,
stated that a higher degree of hyperuniformity (smaller coefficient $A$) is advantageous
for obtaining substantial complete PBGs. In the Appendix, we provide the first
calculations of the coefficient $A$ for two-dimensional quasicrystals, namely, the 5- and
8-fold examples considered here.  The number variance for quasicrystals is greater than
for crystals (for the same density) and less than the variance of disordered
heterostructures.  The results here are consistent with our conjecture in that the width
of the optimal complete band gaps for the photonic quasicrystals lies between the optima
found for crystals and disordered heterostructures.

Although photonic crystals have slightly larger complete band gaps, quasicrystalline PBG
materials offer advantages for many applications.  In the case of quasicrystalline
structures the PBGs are significantly more isotropic, which is advantageous for use as
highly-efficient isotropic thermal radiation sources.\cite{isotropTQC} The properties of
defects and channels useful for controlling the flow of light are different for crystal
and quasicrystal structures. Radiation with frequencies above or below the band edges are
propagating modes that are transmitted through photonic crystals but are more localized
modes in the case of quasiperiodic patterns, which give the former advantages in some
applications, such as light sources.\cite{laser_qc} On the other hand, due to the lack of
translational symmetry, PBG structures constructed around quasiperiodic point patterns can
provide a large number of inequivalent local environments and as such can support a rich
variety of localized modes. These localized modes can be employed to design laser systems
with highly unusual field patterns with possible applications to biological
sensing.\cite{qcSensor}

\begin{acknowledgments}
This work was supported by National Science Foundation under Grant No. DMR-0606415.
\end{acknowledgments}

\appendix*
\label{appendixA}
\section{Asymptotic Number Variance for the 5- and 8-Fold Quasicrystal Point Patterns}

Here we compute the asymptotic number variance for the 5- and 8-fold quasicrystal point
patterns in two dimensions obtained from the Penrose and octagonal tilings. For such point
patterns, the number variance for large windows grows like the perimeter of the window,
i.e.,
\begin{equation}
\langle N^2_R\rangle -\langle N_R\rangle^2= A \frac{R}{D} + {\cal{O}}(1), \qquad R
\rightarrow \infty,
\end{equation}
where $R$ is the window radius and $D$ is the mean-nearest neighbor distance associated
with the point pattern.  In order to compare the dimensionless surface area coefficient
$A$ for different point patterns, the comparison must be made at the same number density
$\rho$ (number of points per unit volume). It was shown in Ref. \onlinecite{hyperunif}
that this amounts to rescaling $A$ such that $A/\sqrt{\phi}$, where
\begin{equation}
\phi=\rho \frac{\pi}{4} D^2
\end{equation}
is an effective packing fraction. Thus, we define the rescaled coefficient $\tilde{A}
\equiv A/\sqrt{\phi}$.  

The rescaled surface area coefficients can be evaluated numerically by employing finite
large quasicrystalline point patterns (of around 15,000 points) and count the number of
points in a circular window of radius $R$. For a fixed radius, the circular window scans
the quasicrystalline domain and the variance of the number of points in the window is
evaluated (the windows employed in our calculations contain from a few to 5000 points and
to calculate the variance we use 1000 configurations with a fixed window size). The
surface area coefficient is then obtained by analyzing the behavior of the number variance
in the limit of large window size.

We find that the rescaled surface area coefficients for the 5- and 8-fold case are given
respectively by $\tilde{A}_{\text{5-fold}}= 0.60052$ and $\tilde{A}_{\text{8-fold}}=
0.59567$.  Our results for these quasicrystal cases should be compared to results for some
periodic point patterns; for the triangular, square, honeycomb and Kagom{\'e} lattices,
$\tilde{A}_{\text{triangular}}= 0.508347$, $\tilde{A}_{\text{square}}= 0.51640$,
$\tilde{A}_{\text{honeycomb}}= 0.56702$ and $\tilde{A}_{\text{Kagome}}= 0.58699$,
respectively. \cite{hyperunif} It is noteworthy that the triangular lattice has been
proven to have the smallest surface area coefficient among all lattices and is thought to
be the global minimum among all point patterns.  Thus, we see that the quasicrystal point
patterns studied here have a high degree of hyperuniformity, even if not as large as that
for the aforementioned crystals. These calculations of the surface area coefficients and
PBGs reported here are consistent with the conjecture made in Ref. \onlinecite{ref1},
namely, that the width of complete PBGs of high-dielectric contrast photonic structures is
correlated with the degree of hyperuniformity of the underlying point pattern.

\end{document}